\documentclass[a4paper]{spie}  
\usepackage[]{graphicx}
\usepackage{amssymb,amstext}
\usepackage{cite}
\usepackage{hyperref}

\def\R23{\mbox{$\rm R_{23}$}}

\title{The MIRC-X 6-telescope imager: Key science drivers, instrument design and operation
 \footnote{Copyright 2018 Society of Photo-Optical Instrumentation Engineers. One print or electronic copy may be made for personal use only. Systematic reproduction and distribution, duplication of any material in this paper for a fee or for commercial purposes, or modification of the content of the paper are prohibited.}
}

\author{Stefan Kraus\supit{a}, 
John D.\ Monnier\supit{b},
Narsireddy Anugu\supit{a},
Jean-Baptiste Le Bouquin\supit{b},\\
Claire L.\ Davies\supit{a},
Jacob Ennis\supit{b},
Aaron Labdon\supit{a},
Cyprien Lanthermann\supit{c},\\
Benjamin Setterholm\supit{b},
Theo ten Brummelaar\supit{d}
\skiplinehalf
\supit{a}University of Exeter, School of Physics and Astronomy, Stocker Road, Exeter, UK;\\  
\supit{b}Department of Astronomy, University of Michigan, 500 Church St., Ann Arbor, USA; \\ 
\supit{c}Institut de Plan\'etologie et d'Astrophysique de Grenoble, Grenoble Universit\'e Alpes/CNRS, Grenoble, France; 
\supit{d}CHARA Array, Mt.\ Wilson Observatory, Mt.\ Wilson, California, USA
}

\authorinfo{Further author information: \\
Send correspondence to S.K., E-mail: s.kraus@exeter.ac.uk, Telephone: +44 1392 724125}

\begin{document} 
\maketitle 

\begin{abstract}

MIRC-X is a new beam combination instrument at the CHARA array that enables 6-telescope interferometric imaging on 
object classes that until now have been out of reach for milliarcsecond-resolution imaging. 
As part of an instrumentation effort lead by the University of Exeter and University of Michigan, we equipped the 
MIRC instrument with an ultra-low read-noise detector system and extended the wavelength range to the J and H-band. 
The first phase of the MIRC-X commissioning was successfully completed in June 2017.
In 2018 we will commission polarisation control to improve the visibility calibration and implement a
'cross-talk resiliant' mode that will minimise visibility cross-talk and enable exoplanet searches using
precision closure phases.
Here we outline our key science drivers and give an overview about our commissioning timeline.
We comment on operational aspects, such as remote observing, and the prospects of co-phased parallel operations with the upcoming MYSTIC combiner.

\end{abstract}



\section{INTRODUCTION}
\label{sec:intro}

MIRC-X is a 6-telescope beam combination instrument installed at the CHARA telescope array,
 which is the world’s highest-resolution imaging facility in infrared light, located at the Mount Wilson Observatory in California. 
The new instrument combines the light from the six 1\,m CHARA telescopes that are spaced up to 330 metres apart, 
making CHARA/MIRC-X ideally suited for imaging stellar surface structures or to image
 the environment around stars with unprecedented resolution.

The instrumentation project is lead by teams at the University of Exeter (UK) and University of Michigan (USA) and builds on the highly 
successful \textit{Michigan Infrared Combiner} (MIRC) instrument that operated between 2005 and 2017 \cite{mon04,mon06,mon10}  and 
provides exciting imaging results, for instance of spotted stellar surfaces (e.g.\ Roettenbacher et al.\cite{Roettenbacher2016}), 
eclipsing binary systems (e.g.\ Kloppenborg et al.\cite{klo10}), and expanding nova shells (e.g.\ Schaefer et al.\cite{sch14}). 

As part of the European Research Council-funded project ``ImagePlanetFormDiscs'', we equipped the instrument with an ultra-low read-noise 
detector system that is based on the electron avalanche photodiode technology.
The eAPD technology offers a truly revolutionary performance compared to the earlier-generation HAWAII or PICNIC
detector arrays by incorporating an electron avalance multiplication stage that amplifies the signal
before the image is read out.  So far, the read noise has been the dominant noise contributor for high-speed infrared detector
required in astronomical interferometry.  

Our work on the detector is complemented by various improvements to the optical design
that aim at achiving additional sensitivity gains and to enable new science applications,
for instance by extending the wavelength coverage to the J-band and enabling higher-precision
measurements.
In order to minimise the engineering risk, we decided to carry out the engineering work in two phases,
where Phase~1 focussed on replacing MIRC's PICNIC detector with the eAPD based detector system 
and with updating and restructuring the software architecture.
Phase~2 will include improvements to the optics that we plan to implement in September 2018.

Here, we provide a project-level overview about the key science drivers of the project (section~\ref{sec:science}),
the work that was completed in Phase~1 (section~\ref{sec:phase1}), and the planned improvements for
Phase~2 (section~\ref{sec:phase2}).
Finally, we comment on the operational modes that we plan to implement for MIRC-X and first on-sky results
(section~\ref{sec:operation}).

In parallel to MIRC-X, members of our team are also developing the MYSTIC beam combiner 
that will operate in the K-band (Monnier et al.\cite{mon18}).
While both are separate projects, we try to utilise synergies between the two projects, both on the optical design and on the control software.
This approach also allows us to envision specialised operational modes for fringe detection and fringe tracking,
as we will discuss in section~\ref{sec:MYSTICop}.

\section{KEY SCIENCE OBJECTIVES}
\label{sec:science}

The primary science drive for the MIRC-X project is to enable efficient 6-telescope imaging of  the inner regions of protoplanetary disks.
However, whenever possible we optimised our design also for various other potential high-impact science cases.
Below, we list these science objectives in order of priority.

\subsection{Protoplanetary disk structure \& disk physics}

Studying the structure of the disks around young stellar objects, such as low-mass T\,Tauri stars and intermediate-mass Herbig Ae/Be stars,
is crucial to advance our understanding of the star and planet formation process.
The infrared excess emission that is observed towards these objects is believed to trace predominately the emission from hot dust grains.
However, there is still some fundamental uncertainty how the dust is arranged, e.g.\ whether it forms a sharp, 
vertically extended ``puffed-up'' inner dust rim near the dust sublimation radius or a more continuous curved rim (e.g.\ Dullemond \& Monnier\cite{dul10}).
Furthermore, it is believed that planets form in the inner regions of these disks and that these bodies
will dynamically shape their natal disk environment, for instance by opening gaps or triggering asymmetric structures
that might orbit the star on the dynamical timescale of a few months.
Therefore, the primary objective of MIRC-X is to image these disk asymmetries and to trace their motion \& evolution.
This requires optimising the sensitivity of the instrument, which we aim to achieve by commissioning the ultra-low read-noise SAPHIRA-based detector system.

\subsection{High-precision astrometry for exoplanet detection}

A potential high-impact science case for MIRC-X is exoplanet detection, either through precision closure phases
or differential single-field astrometry in close binary systems.

Zhao et al.\cite{zha11} investigated the feasibility of detecting some of the brightest Hot Jupiter systems with CHARA and predicts a
closure phase signal of $\sim 0.05^{\circ}$ for Ups And b.  This closure phase precision might be within reach
for MIRC-X, if the fringe cross-talk can be minised and all systematics can be well-characterised and corrected.

An alternative approach is to search for S-type exoplanets in close binary systems that orbit one of the stellar components.
The gravitational influence of the planet causes a 'wobble' in the differential astrometry of the binary system.
Gardner et al.\cite{gar18} presented MIRC observations of the close binary system $\delta$\,Del and demonstrated
that it is possible to achieve a precision in the relative binary astrometry of a few micro-arcsecond.
This precision should enable to detect a Jupiter-mass S-type exoplanets at a separation of $\sim 2$ astronomical units
or a 4 Jupiter-mass planet at 0.5\,au separation.  

\begin{figure}[tb]
\centering
\hspace{14mm}\includegraphics[width=0.4\textwidth]{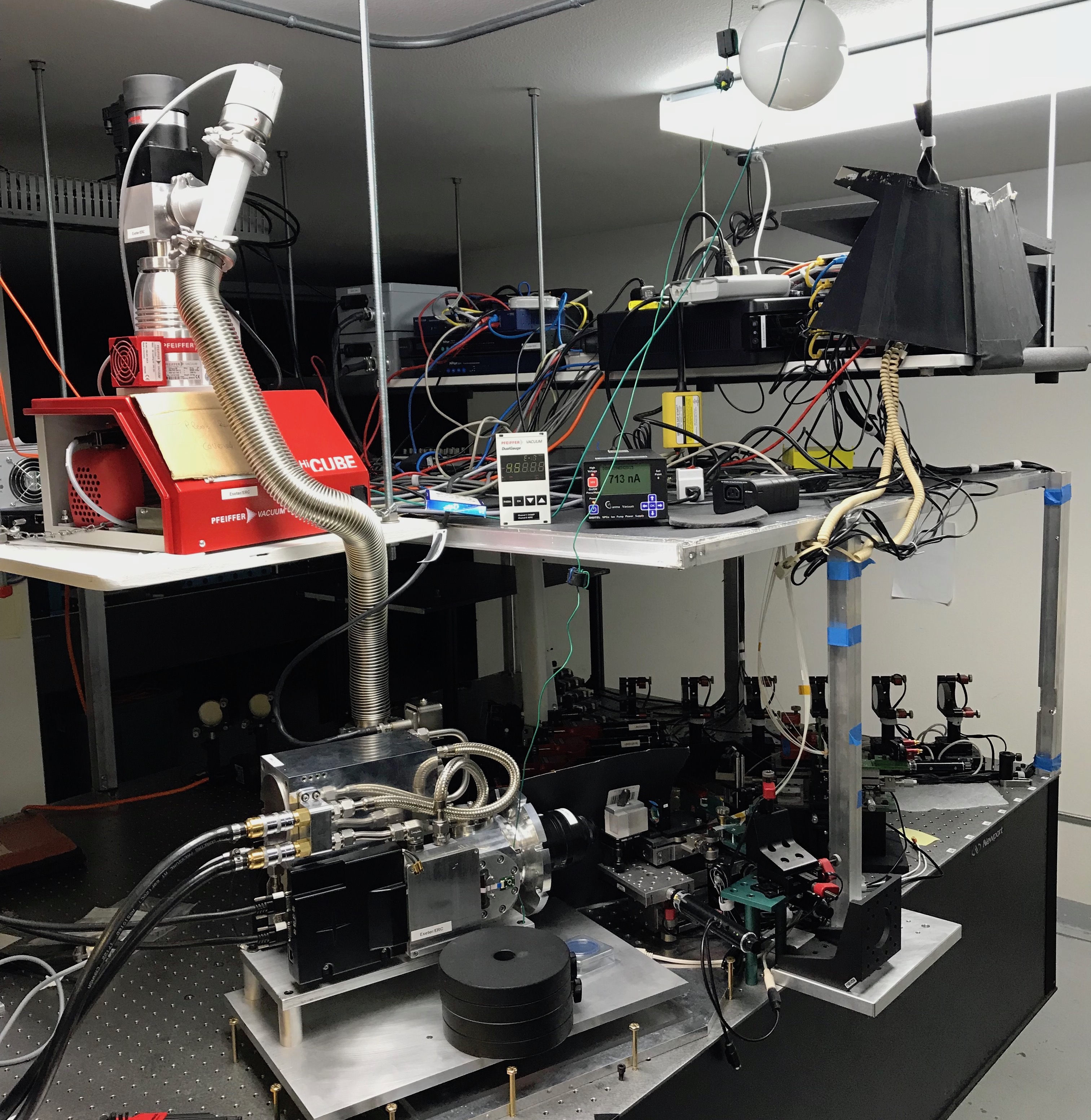} 
\includegraphics[width=0.48\textwidth]{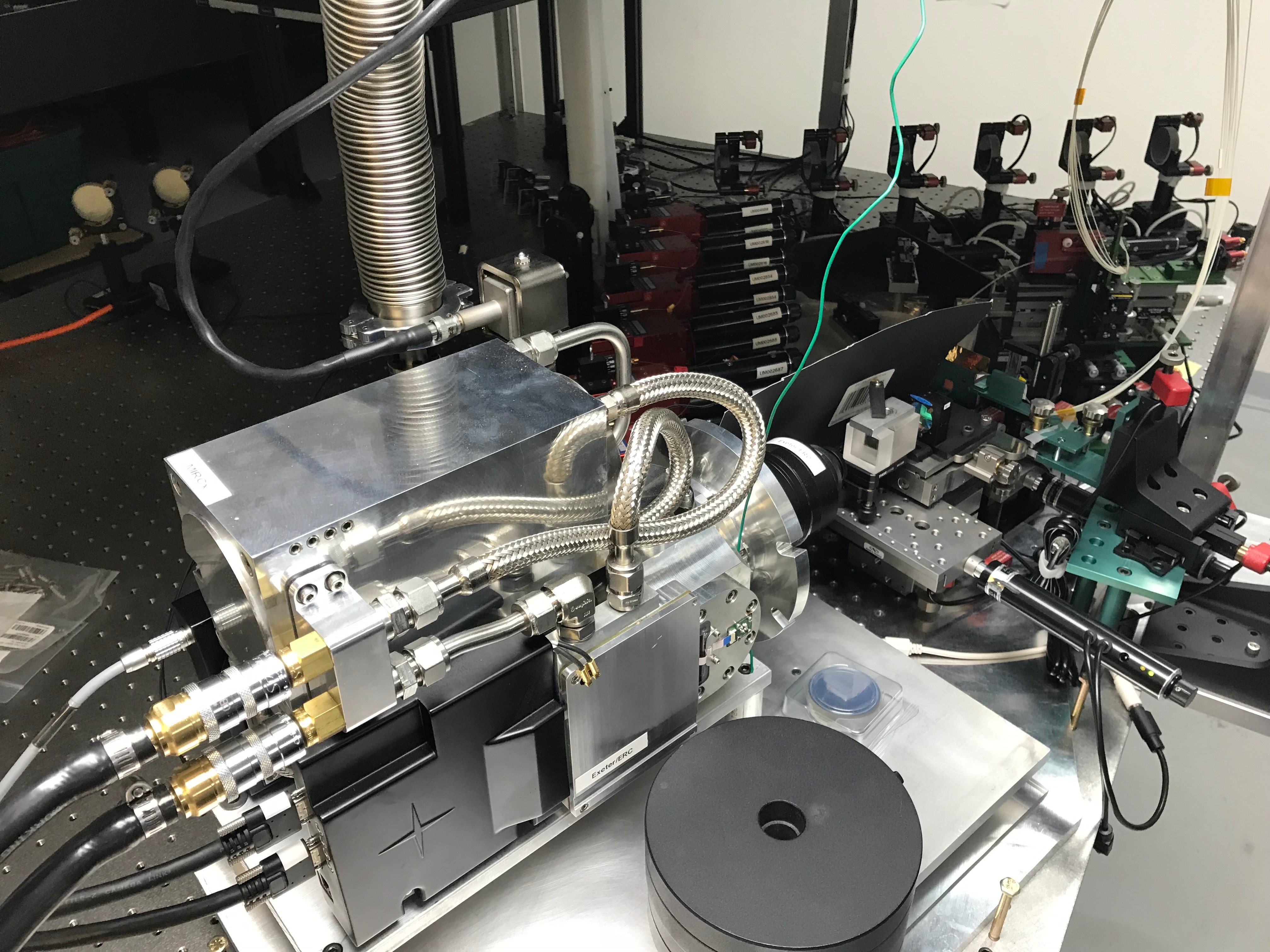} \\
\includegraphics[width=0.48\textwidth]{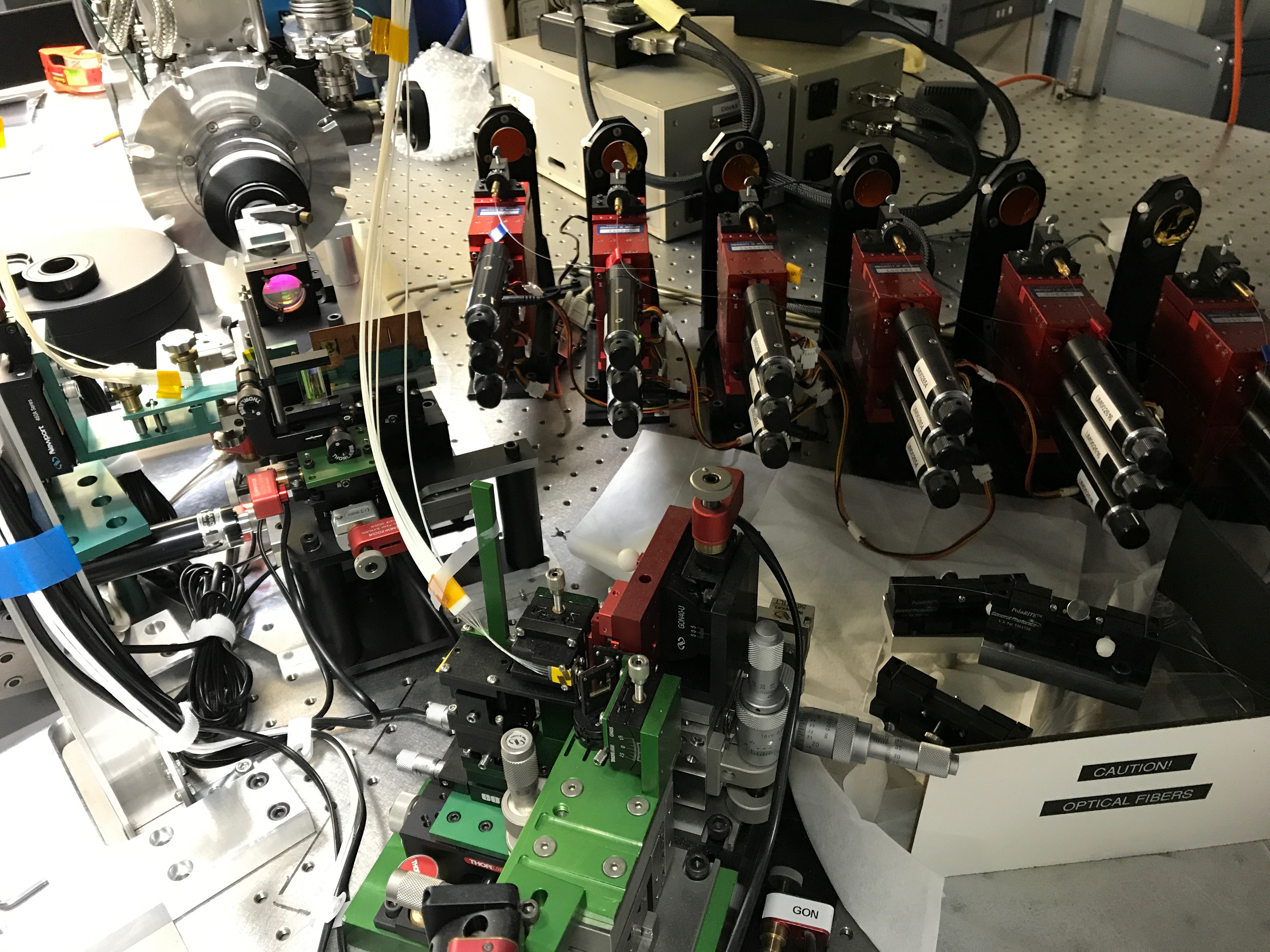} 
\includegraphics[width=0.48\textwidth]{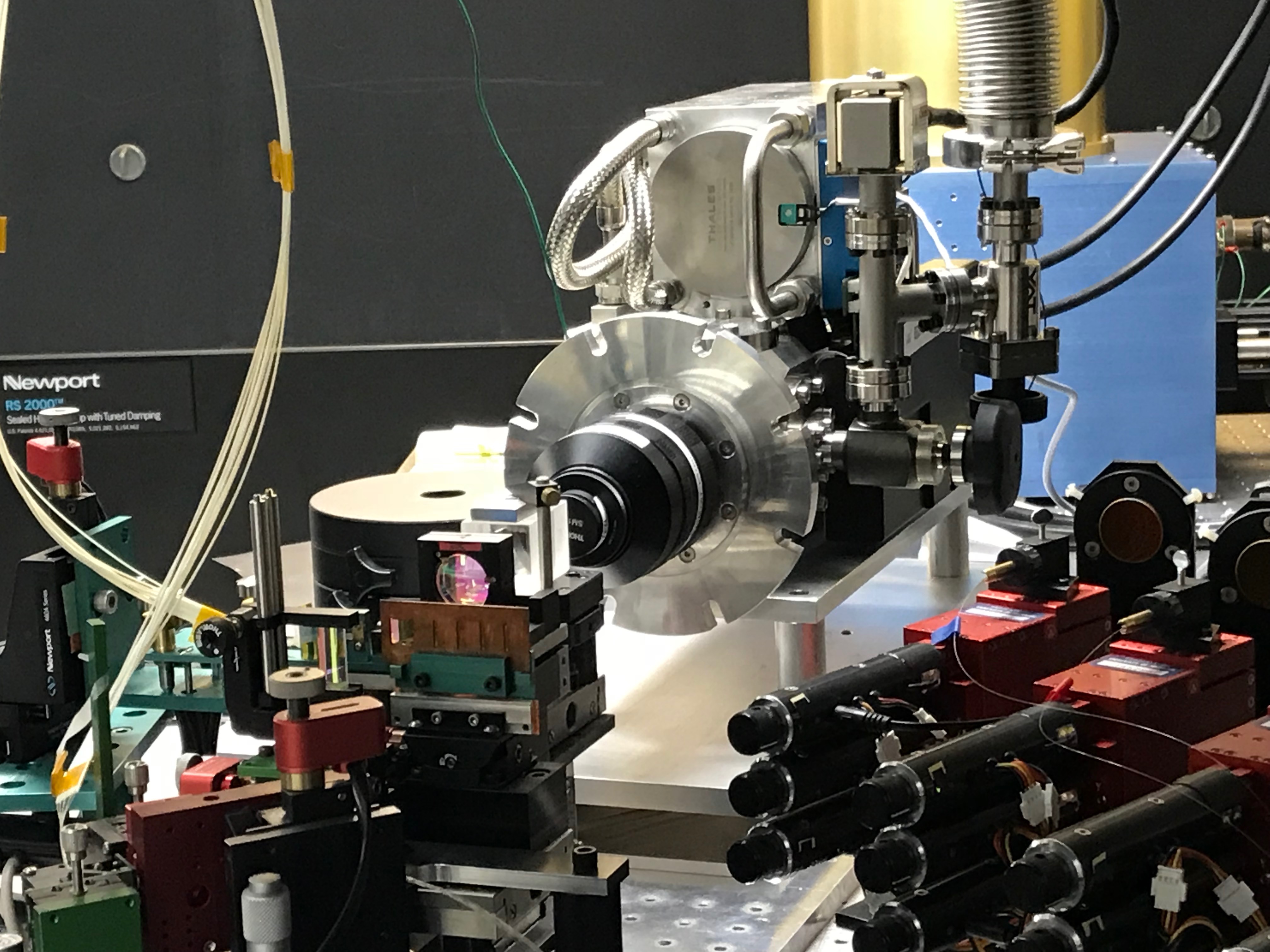} 
\caption{ MIRC-X system after completion of the Phase~1 work, which included the installation of the C-RED One camera system and associated components to maintain vacuum (turbo pump, ion pump) and cooling. }
\label{fig:photo}
\end{figure}

\subsection{Polarimetric Interferometry for scattered light imaging}

We implement polarisation control in MIRC-X in order to improve the visibility calibration accuracy,
but note that the implemented hardware could potentially also be used for measuring polarised light of astrophysical origin.
Splitting the two polarisation states could enable imaging in Stoke U and Q parameters, which could provide exciting new
information about the dust scattering properties on astronomical unit scales, for different interesting astrophysical object classes.
Commissioning such a science polarimetry mode is out of the scope of our present instrumentation plan,
but we anticipate pursuing this science in the intermediate term.

\subsection{Spectro-interferometry in J and H-band lines}

Interferometry with high spectral resolution in spectral lines can provide direct constraints on the spatial distribution of the line-emitting gas.
This technique has already been used successfully in the $K$-band for measuring the origin of the Br$\gamma$ line emission 
(e.g.\ Kurosawa et al.\cite{kur16}), Pfund series line emission (e.g.\ Kraus et al.\cite{kra12c}), and the CO overtone bandhead emission (e.g.\ Eisner et al.\cite{eis14}).
These spectral lines trace fundamental processes, such as magnetospheric accretion, stellar winds, and mass loss.
We plan to implement spectral resolution modes with R=1170 (for J-band) and R=1035 (for H-band) that could enable detections in these lines,
although an efficient phase-tracking mode with MYSTIC will be required in order to enable the long integration times that are
necessary to accummulate sufficient signal-to-noise (SNR) for observations at this spectral resolution.

\section{PHASE 1: DETECTOR UPGRADE AND OPERATIONAL IMPROVEMENTS}
\label{sec:phase1}

\subsection{eAPD detector implementation}

\begin{figure}[tb]
\centering
\includegraphics[width=0.55\textwidth]{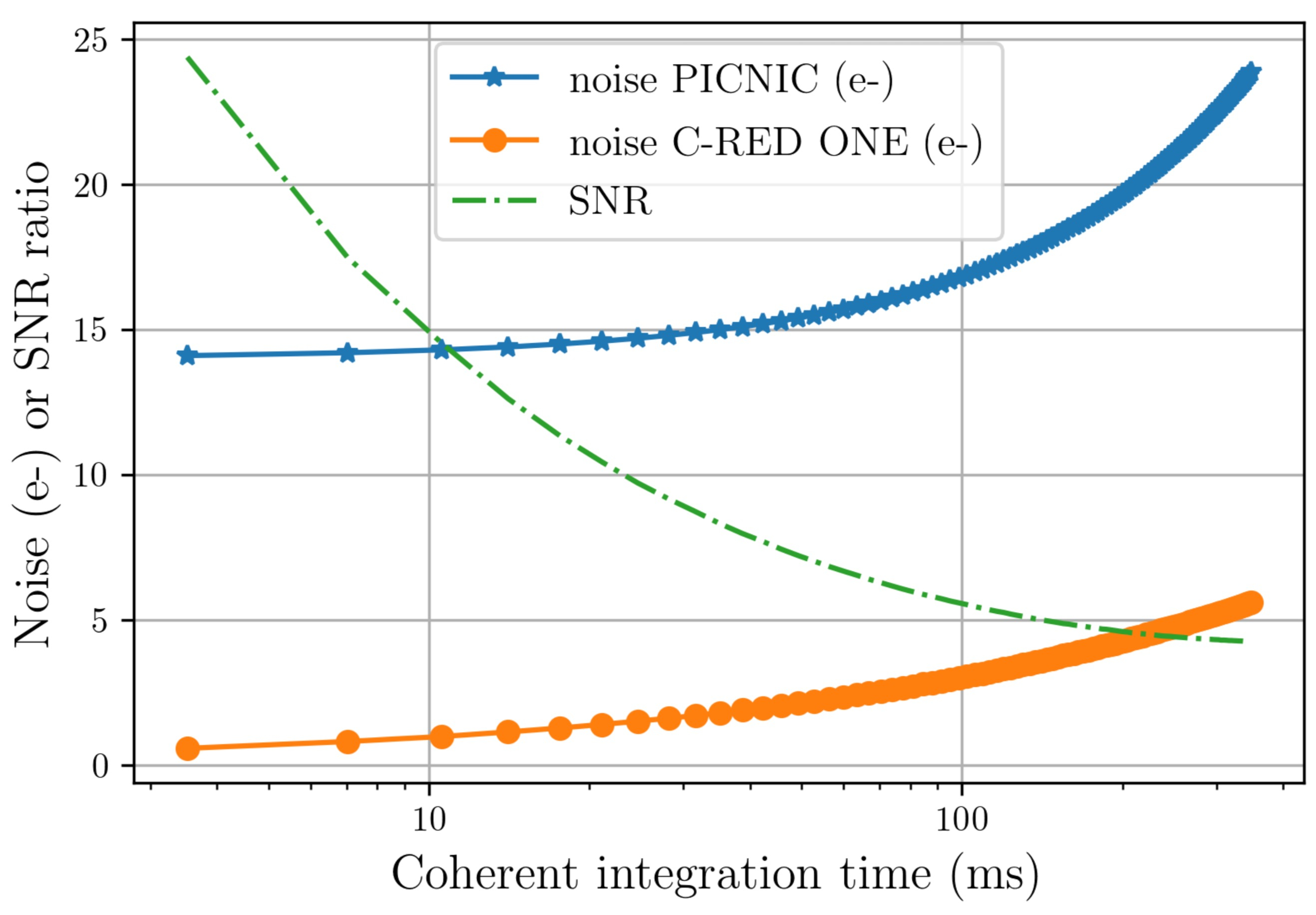} 
\caption{ Computed total noise for the MIRC PICNIC detector (blue line) and the MIRC-X C-RED One (orange line), plotted as function of coherent integration time.
The green dash-dotted curve shows the corresponding SNR improvement. }
\label{fig:SNRimprov}
\end{figure}

\begin{figure}[tb]
\centering
\includegraphics[width=0.8\textwidth]{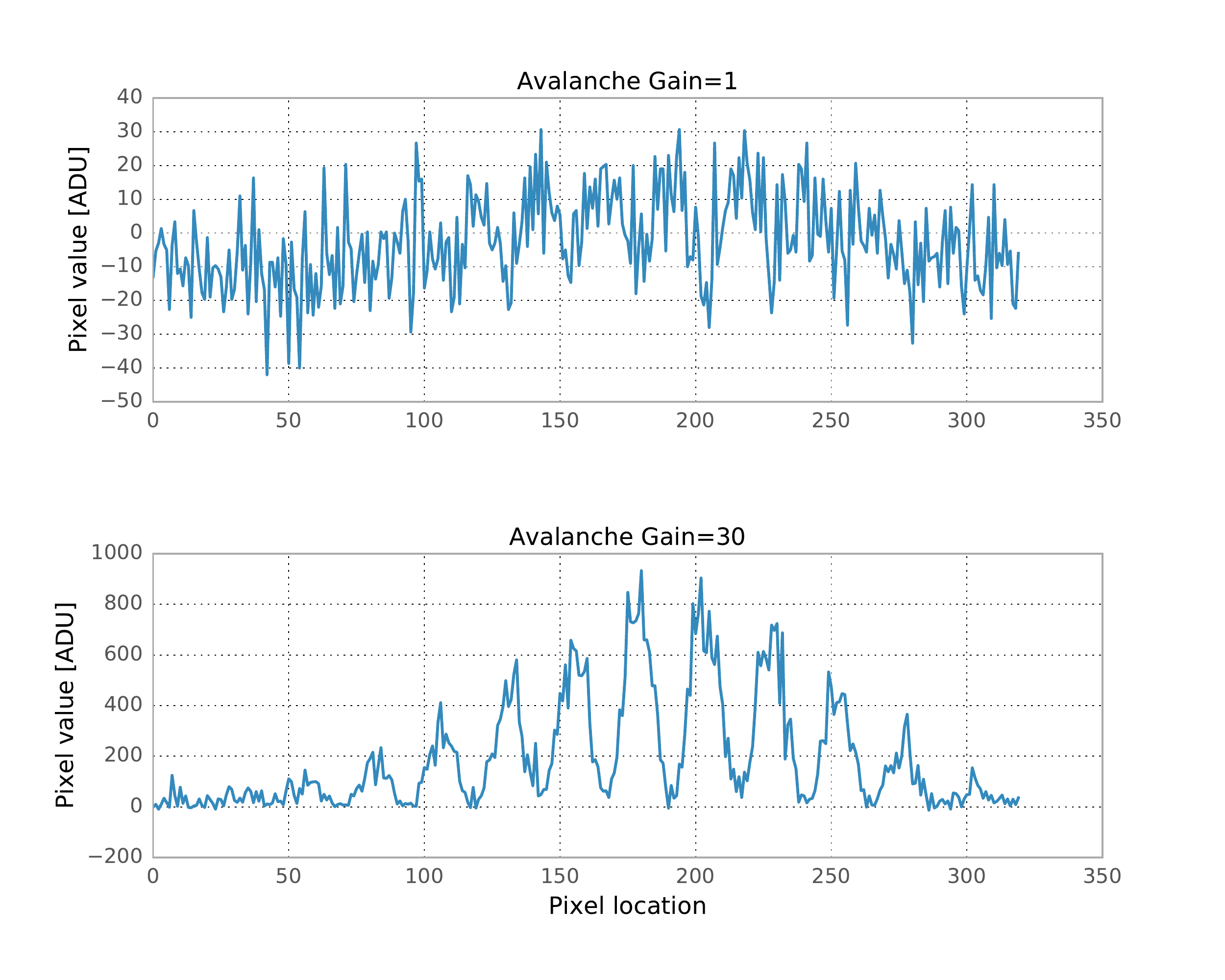} 
\caption{ Fringes obtained with the CHARA laboratory light source, observed at gain 1 (top) and gain 30 (bottom). }
\label{fig:gain}
\end{figure}

The focus of the Phase 1 work has been on the characterisation, integration, and on-sky commissioning of the
eAPD based detector system, while keeping most of the existing MIRC optics in place.
For this purpose, we purchased from the French company First Light Imaging (FLI) SAS 
a C-RED One camera system\cite{credone2016} that incorporates the SAPHIRA detector\cite{finger2014} developed by the UK company Selex/Leonardo.
This detector offers $320\times256$ pixel with an avalance amplification gain up to 500. 
Being the first C-RED One  customer, we also had a welcome opportunity to work with FLI in order to optimise
the camera performance, for instance by suggesting the implementation of a custom Fowler sampling read-out mode
that helped to push down the noise.
The camera is operating at 80\,K using a pulsed tube cryocooler, where excess heat is removed using a closed-cycle water cooling system.
In order to maintain the vacuum in the camera dewar, we initially operated the system with a vacuum turbo pump.
Given the turbo pump might introduce mechanical vibrations, we then added an ion pump that allows us to maintain 
the camera vacuum without running the turbo pump (see Fig.~\ref{fig:photo}).

We could confirm that the detector provides an order-of-magnitude improvement in sensitivity (reaching a sub-electron total noise for integrations below $\sim 50$\,Hz; Lanthermann et al.\cite{lan18}), 
in particular when using short coherent integration times. 
For instance, for short coherent integrations of 5\,ms, as might be used on bright stars and in poor seeing conditions,
we compute an improvement in interferogram SNR of factor 20 for the MIRC-X detector compared to MIRC.
For long coherent integrations of 50\,ms, as might be necessary for the observation of faint target stars,
the sensitivity gain is less pronounced (factor $\sim 7$, or about 2\,magnitudes), as the thermal background 
becomes the dominant noise source in this regime (see Fig.~\ref{fig:SNRimprov}).
A detailed report on our in-lab characterisation of the camera and of the mathematical framework 
that we use for measuring the detector performance, can be found in Lanthermann et al.\cite{lan18} 
and Anugu et al.\cite{anu18}.

Figure~\ref{fig:gain} illustrates how the avalance gain amplifies the signal (here as measured on an artifical light source in the laboratory) and can pull a signal out of the noise.
We also observe some sensitivity gain on sky, although the improvement was lower than measured on laboratory fringes.
Likely reasons for this is the present oversampling of the fringes, where the existing optics is still optimised for
the pixel-scale of the old PICNIC detector (40\,$\mu$m/pixel).
This will be adjusted in Phase~2 in order to match the pixel-scale of the new detector (25\,$\mu$m/pixel).
Also, during our 2017 runs, we often observed a poor beam quality from the telescopes, which resulted in
a poor injection into MIRC-X's single-mode fibers.
In order to improve the beam quality, we used CHARA's new lab-AO system in order to apply a static correction
and to better focus the beams. We expect further improvements in this respect with the arrival of the deformable
mirrors and the deployment of the full adaptive optics system in late 2018.

\begin{figure}[tb]
\centering
\includegraphics[width=1.0\textwidth]{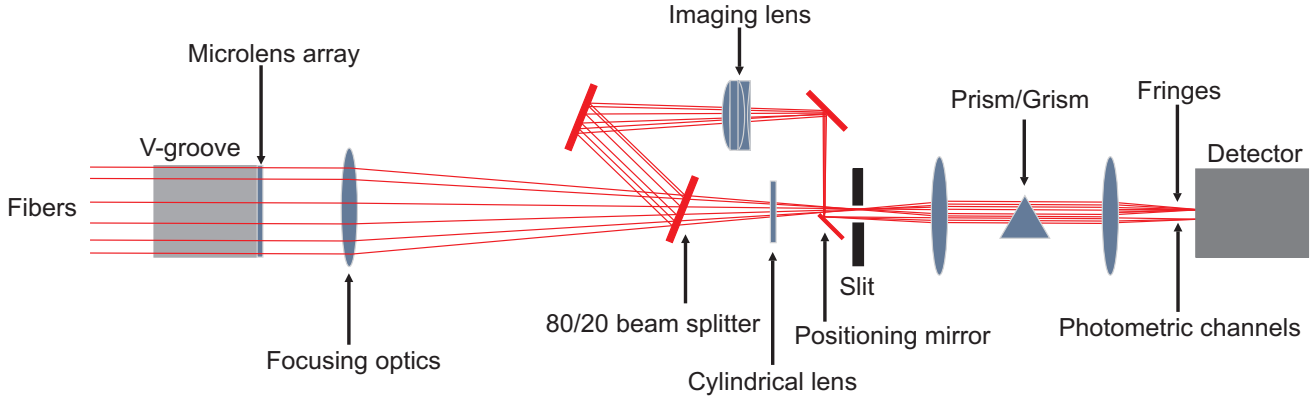} 
\caption{ Sketch of the optical path for the MIRC-X system (see Monnier et al.\cite{mon06} to compare with MIRC design). The sketch is not to scale.}
\label{fig:design}
\end{figure}

\subsection{Computer hardware and software architecture}

As part of Phase~1, we also installed a new rack instrument computer that is able to process and 
store the large data stream volume provided by the C-RED One camera, which amounts to about 
540\,MB/s ($320 \times 256\,\mathrm{pixels}^2 \times 16\,\mathrm{bit} \times 3500\,\mathrm{frames/s}$),
or nearly 2\,TB per hour.
This data stream is facilitated through a dual CameraLink connection, where the electronic signal is 
converted into an optical signal near the camera, then transmitted from the optical laboratory to the
computer room through an optical fiber, and converted back into an electronic signal and digitised with a frame grabber.
To enable the required high data throughput, we first record the data on a SSD disk,
before transferring it to several HDD disks at the end of each observing night.

Besides the data recording, the instrument computer also carries out the real-time data processing
that is needed for fringe search and group delay tracking.
As outlined in Anugu et al.\cite{anu18}, we implemented all real-time operation functionality in a C-based MIRC-X server
that communicates through a server-client approach with the user-facing GUIs.
The GUIs utilise the GTK graphics toolkit and cover a wide range of functionality, including the camera setup (``ircam''),
instrument setup (``parameters''), flux-injection optimisation into the fibers (``fiberexplorer''), 
the fringe search and control of the group delay tracking (``gdt''), the real-time display of the data (``rtd''), and 
the data recording (``sequencer'').
All GUIs have been structured efficiently so that they fit on a single screen.
Details about the software architecture and the functionality of the new GUIs can be found in Anugu et al.\cite{anu18}.

\section{PHASE 2: OPTICS RE-DESIGN }
\label{sec:phase2}

\begin{figure}[tb]
\centering
\includegraphics[width=0.8\textwidth]{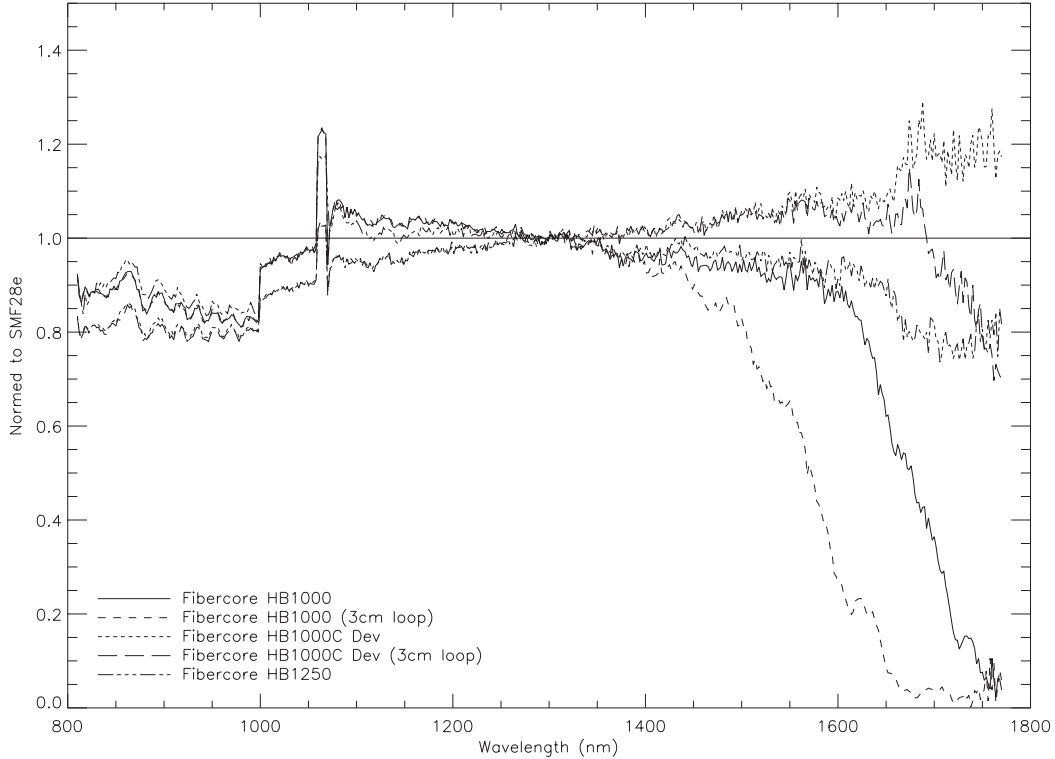} 
\caption{ Spectra measured for three polarisation-maintaining single-mode fibers that we considered for MIRC-X.
All spectra are compared to reference fiber SMF28e and normalized to 1200-1250\,nm. }
\label{fig:fiber1}
\end{figure}

\begin{figure}[tb]
\centering
\includegraphics[width=0.8\textwidth]{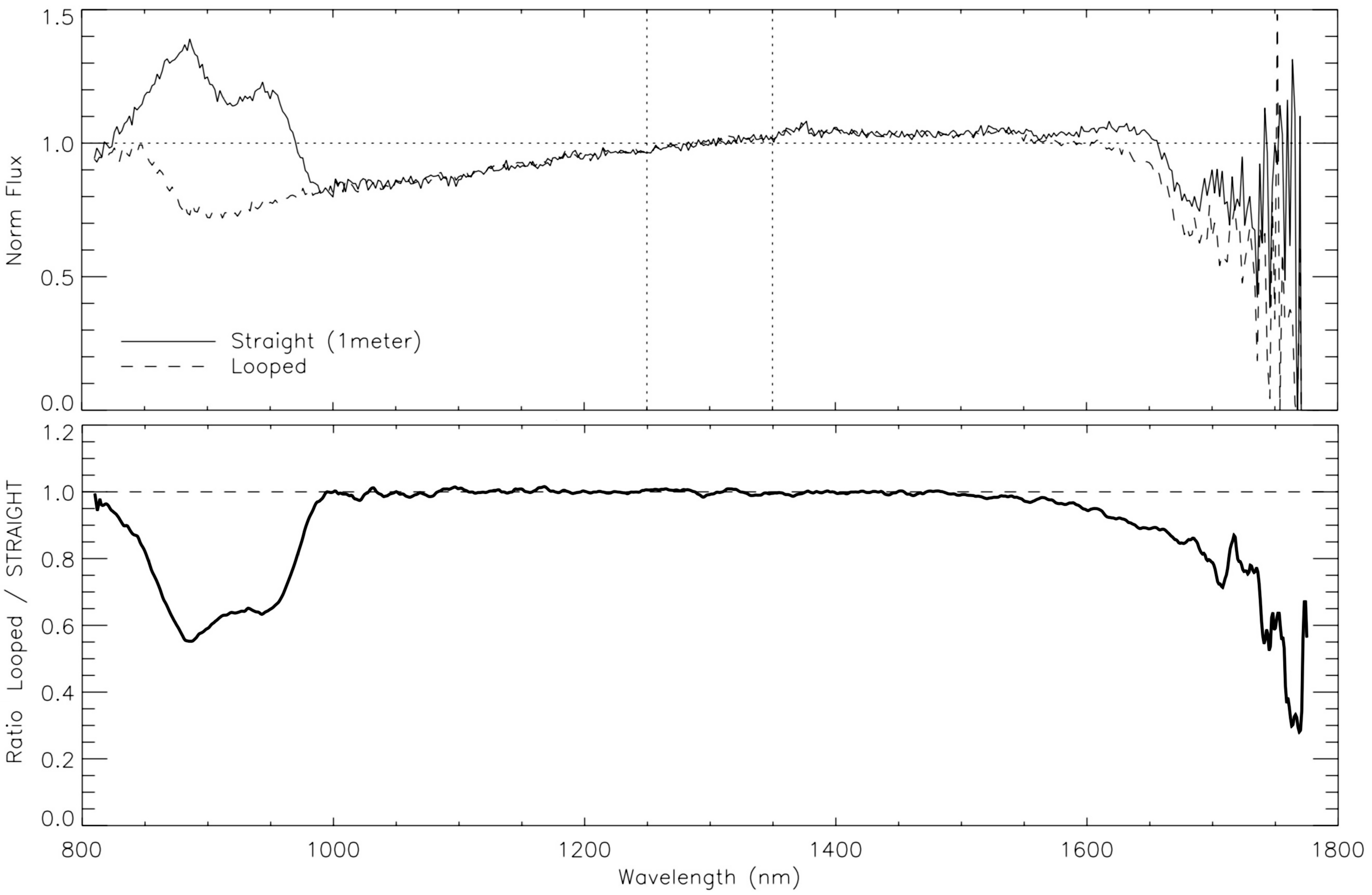} 
\caption{ Bending losses characterisation of the Fibercore HB1000C(6/125)\,DEV fiber that we use for MIRC-X. }
\label{fig:fiber2}
\end{figure}

\subsection{Optical layout}

As part of Phase 2, we will rearrange the optics on the MIRC optical table, both to clear space
for accommodating the MYSTIC instrument and to implement substantial optical improvements.
Motorized pick-off mirrors will reflect the light from the CHARA beams towards fold-mirrors 
that redirect the light towards off-axis parabolas that focus the light into polarisation-maintaining single-mode fibers.  
The light from the 6 fibers is arranged with a non-redundant spacing on a V-groove
and then collimated with a microlens array.
The light then passes through a focusing lens and the photometric signal is extracted with 
a 80/20 beam splitter.
A slit-forming cylindrical lens compresses the fringe pattern in the spectral direction to define the 
input slit of the spectrograph.
The light from the individual beams is then re-imaged into the slit.
Both the interferometric and photometric channel pass through the spectrograph
before being registered on the C-RED One detector.

\subsection{Redesigned photometric channels}

We implement a novel concept for recording the photometric channels and depart from
the MIRC design that re-injected the light into multi-mode fibers after the microlens-array and
rearranged their output on the slit.  The old design worked reliably, but was also very sensitive
to alignment problems, which often limited the accuracy of the recorded relative photometric information.
Furthermore, the $45^{\circ}$ refection of the beam-splitter introduced possible polarisation effects. 

Therefore, we implement a new scheme, where the light is split off after the microlens array
at a right angle with a weakly-polarizing beamsplitter.
The light of the individual six photometric channels passed then through an imaging lens and is reimaged 
with a positioning mirror alongside the interferometric beam (Fig.~\ref{fig:design}).
We anticipate that this design will be less sensitive to alignment errors and could result in a better transmission
for the photometric signal.

\subsection{Polarisation-maintaining fibers and J/H band operation}

In order to optimise the H-band transmission, and to enable J-band observations,
we selected new single-mode fibers for spatial filtering.
At the same time, the fibers need to be polarisation-maintaining in order to 
enable split polarisation observations.

We searched a wide range of vendors in order to find a suitable single-mode fiber
that is polarisation maintaining, has good transmission over the J and H band,
low bending losses, and matches the required numerical aperture (NA=0.18).
After identifying the most promising candidate fibers from different vendors,
we measured the transmission spectra for 14 candidate fibers, both for a straight 
sample (1\,m length) and a strong bend sample (3\,cm loop).
In the bend case, strong guiding losses occur that allow us to
estimate the cut-off wavelength for the probed fiber.

Figure~\ref{fig:fiber1} shows the spectra for the three most promising 
fibers that we identified, all from the company Fibercore, namely the HB1000, HB1000C(6/125)\,DEV, and HB1250.
Our measurements show that the HB1000 shows strong bend guiding losses above $\sim 1550$\,nm
and was excluded for this reason.  
Both the HB1250 and the HB1000C(6/125)\,DEV provide a good match for our requirements.
The HB1250 showed 20\% bend loss beyond 1600\,nm with an extreme, 3\,cm loop bend.
By comparison, the HB1000C(6/125)\,DEV (Fig.~\ref{fig:fiber2}) showed no bending loss 
for minor bends and just 20\% loss at 1.8 microns with the extreme bend.
Therefore, we decided on the HB1000C(6/125)\,DEV (Fig.~\ref{fig:fiber2}), after confirming
that the fiber is single-mode down to 1100\,nm, which is very close to the 1080\,nm HeI line that we aim to target.
The vendor noted that the favourable properties of this fibers are due to the low boron level and Bow Ties that are
located very far (about 20\,$\mu$m) from the fiber core. 
We look forward to confirming these properties in daily instrument operation and on-sky.

One of the challenges associated with observing in J-band at CHARA is contamination from the 
metrology laser that is used for measuring the position of the delay line carts.
This laser transmits at 1.139\,$\mu$m and could strongly saturate the spectral
channels in the center of the J-band.
Therefore, we use a notch filter that will be included for J-band observations
to supresses the laser line.

\subsection{Fringe cross-talk resilient mode}

\begin{figure}[tb]
\centering
\includegraphics[width=0.8\textwidth]{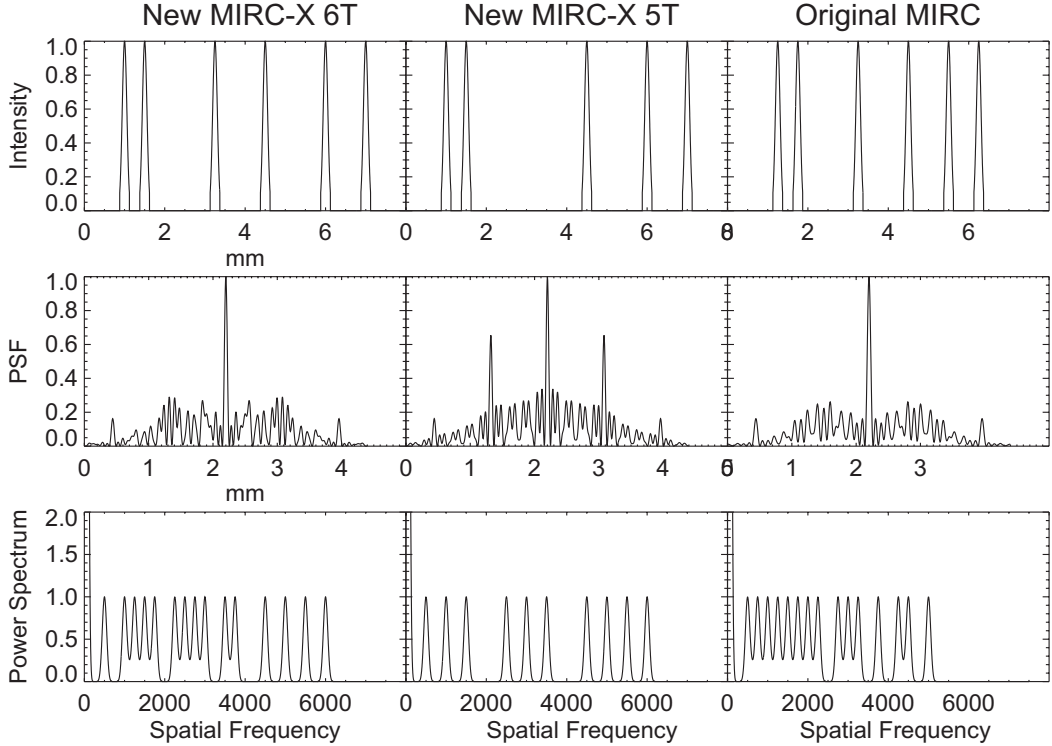} 
\caption{ Fiber arrangement in MIRC-X 6T mode (left) and the 5T(a) 'cross-talk resiliant' mode (middle, with beam~3 shuttered out). 
The panel on the right shows the old fiber arrangement in the MIRC V-groove. }
\label{fig:fiberarr}
\end{figure}

While keeping the multi-axial combination scheme, we decided to optimise the arrangement of the 
fibers on the V-groove array.
MIRC's fiber arrangement resulted in non-redundant fringe frequencies in the power spectrum (Fig.~\ref{fig:fiberarr}, right), 
but still allowed minimal cross-talk between neighbouring fringe peaks on the level of $\sim 1$\%.
For the visibility amplitude estimation, the cross-talk can be corrected reasonably well as part of the data reduction process.
However, such a correction is more difficult to facilitate for closure phase estimation, 
which might have affected MIRC's closure phase precision and limited the success of earlier exoplanet search missions with the instrument.

Therefore, for MIRC-X we re-designed the V-groove fiber arrangement and found a fiber configuration
that allows one to switch from a standard 6T imaging mode (Fig.~\ref{fig:fiberarr}, left) to a 5T 'cross-talk resiliant' mode (Fig.~\ref{fig:fiberarr}, middle)
by shuttering out the light from one telescope.  
Our fiber arrangement permits two different cross-talk resilient configurations,
where either beam 3 or, alternatively, beam 4, is dropped (referred to as 5T(a) and 5T(b) in Table~\ref{tab:modes}).
It is possible to switch very rapidly between these two configurations, therefore allowing to record data 
with minimal overhead in both configurations to improve the uv-coverage.

\subsection{Spectral modes}

\begin{table}[t]
\caption{MIRC-X operating modes}
\label{tab:modes}
\centering
\vspace{0.1cm}
\begin{tabular}{c c c }
\hline\hline
 Mode                                & Spectral band                       & Number of telescopes \\
 \hline
High sensitivity imaging  &  H                 & 6T\\
High-precision astrometry &  H             & 5T(a), without beam 3\\
High-precision astrometry  &  H             & 5T(b) , without beam 4\\
Multi-band                            &  J+H            & 4T\\
Spectro-interferometry      &  H (R=1035)      & 6T\\
Spectro-interferometry      &  J  (R=1170)       & 4T\\
 \hline
\end{tabular}
\end{table}

To match the requirements posed by different science target stars,
we plan to offer different spectral dispersing elements, including three prisms
and up to four grisms.
The prisms will offer spectral resolution R=22, R=50, and R=102.
The R=22 prisms will offer the highest sensitivity but also the
smallest field-of-view.
In order to avoid bandwidth smearing effects on objects with
more extended structures, such as binaries, we 
require a larger
field-of-view that can be achieved with the R=50 or R=102 prism
or the R=182 and R=625 grism.
These dispersing elements still allow recording the full J and H band.

Two other planned grisms will offer high spectral resolution,
suitable for observations in spectral lines, but cover then only one spectral band at a time.
We anticipate procuring grisms with R=1035 for H-band and
R=1170 for J-band.

\subsection{Birefringence correction and polarisation control}

One of the most-difficult-to-control effects that can lower the fringe contrast in optical interferometers
is birefringence between the different light paths that are combined.
Any imbalances will cause a relative shift between the orthogonal polarisation states and a
loss of coherence in the polarization-insensitive interferogram pattern.
Therefore, we decided to follow the birefringence correction method outlined in 
Lazareff et al.\cite{lazareff2012} and to insert Lithium Niobate plates in the optical path
that introduce an adjustable phase-shift between the vertical and horizontal polarisation axes
for each beam before the light is injected into the single-mode fibers.  
The phase shift can be adjusted by rotating the Lithium Niobate plates using
motorized rotary stages.
The rotary stage position will be optimised at the beginning of each observing night
using a new alignment GUI that will measure the fringe contrast as function of rotation angle
for each beam separately and then moves the stages to the position to achieve maximum contrast.

Besides the Lithium Niobate plates, we also plan to implement a Wollaston that will
allow us to split the two polarisation states and to measure the interferometric observables 
in the Q and U states directly. This will provide an alternative approach for calibrating polarisation effects
in the measured visibility. In the medium term, this approach could also allow us to directly 
measure astrophysical polarisation signals that might be caused by dust scattering,
although further resources will be needed to implement a calibration strategy and to
fully commission this mode.

\section{MIRC-X OPERATION}
\label{sec:operation}

\subsection{Commissioning and on-site operation}

\begin{figure}[tb]
\centering
\includegraphics[width=0.8\textwidth]{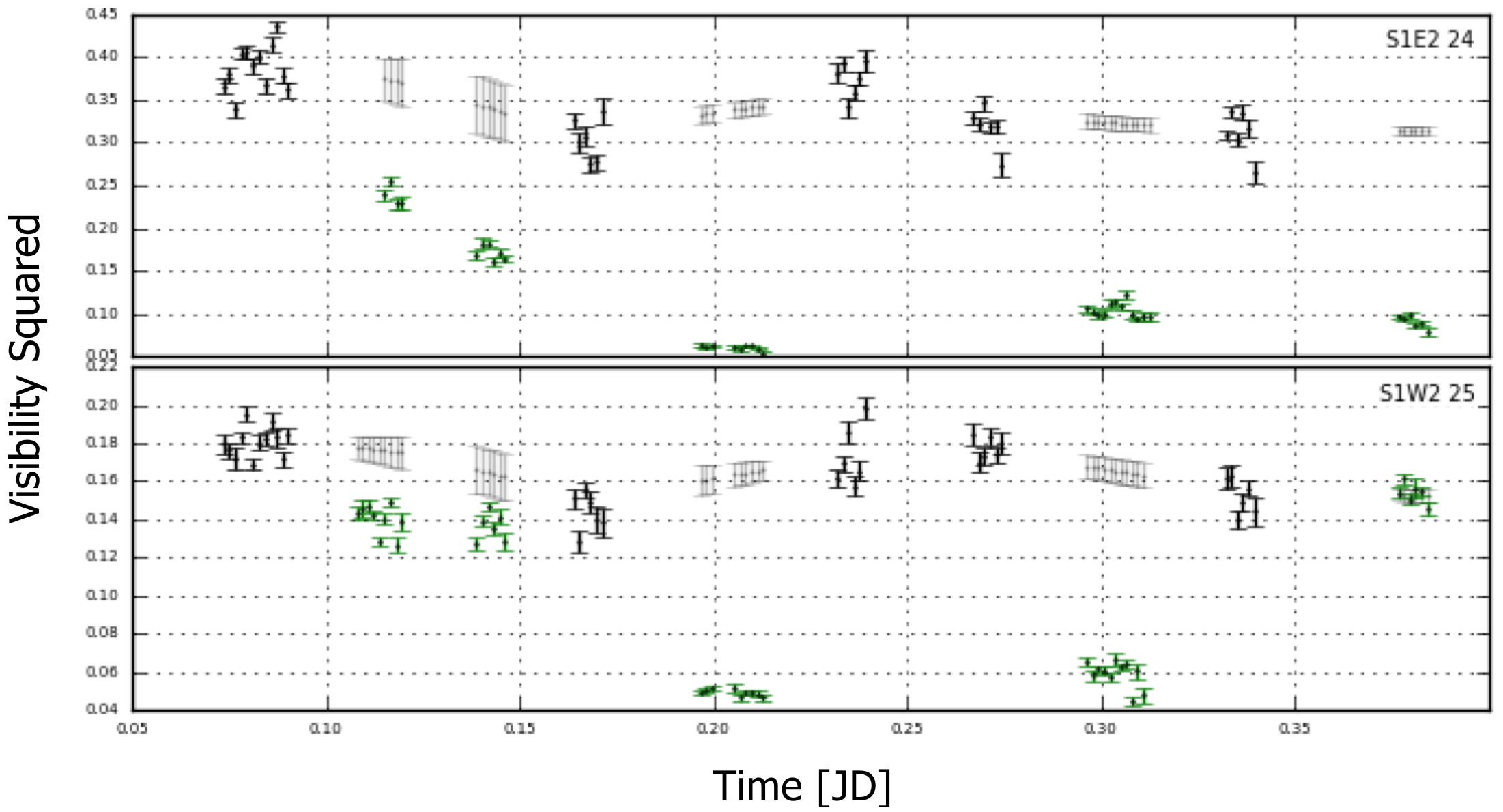} 
\caption{ Typical visibility transfer function, with two calibrators (black and grey data points) and a resolved science star (greeen data points), for two of the 15 baselines.  }
\label{fig:transff}
\end{figure}

\begin{figure}[tb]
\centering
\includegraphics[width=0.8\textwidth]{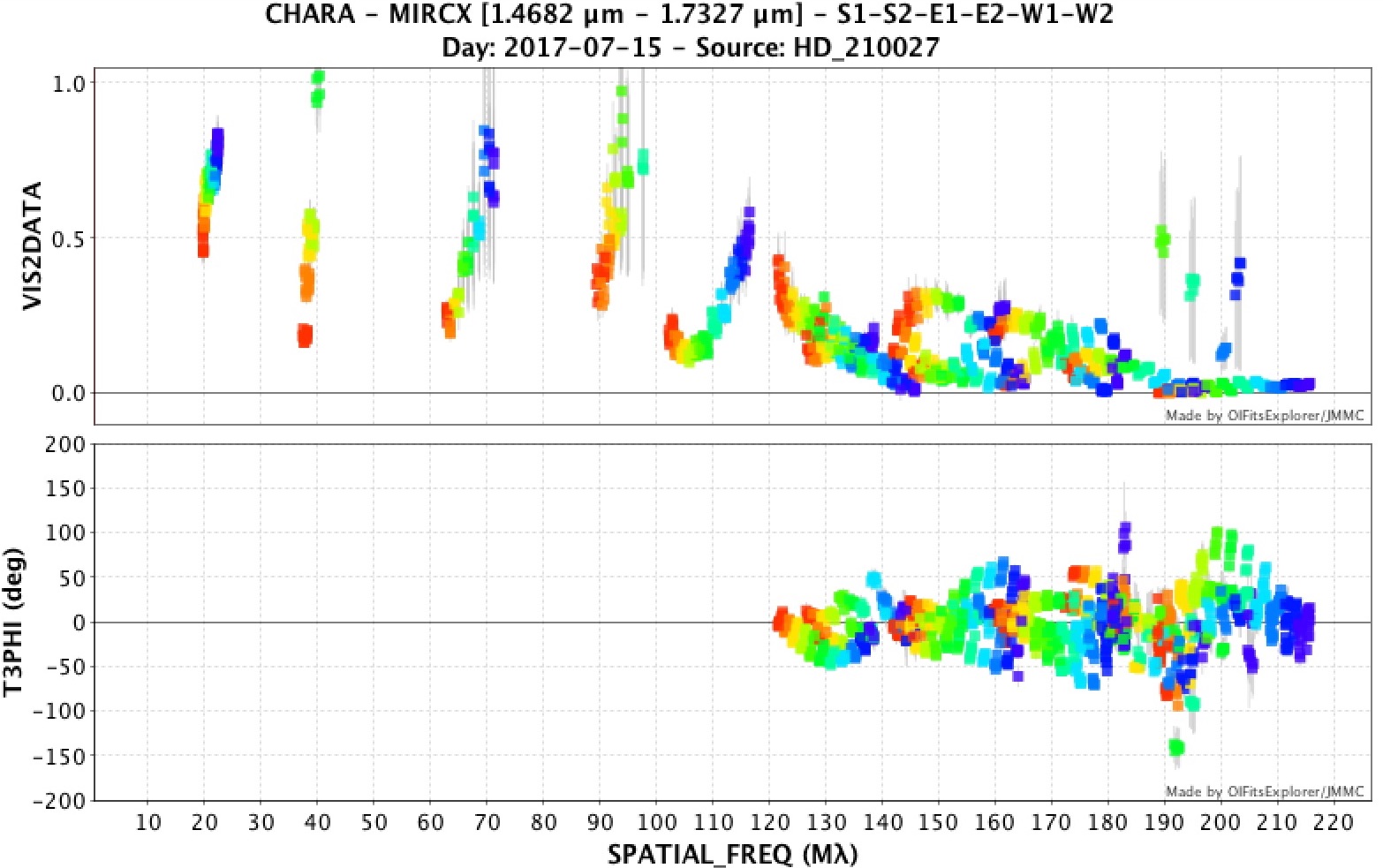} 
\caption{ Visibility and closure phase profile measured on the $\iota$\,Peg binary star. }
\label{fig:iotapeg}
\end{figure}

\begin{figure}[tb]
\centering
\includegraphics[width=0.8\textwidth]{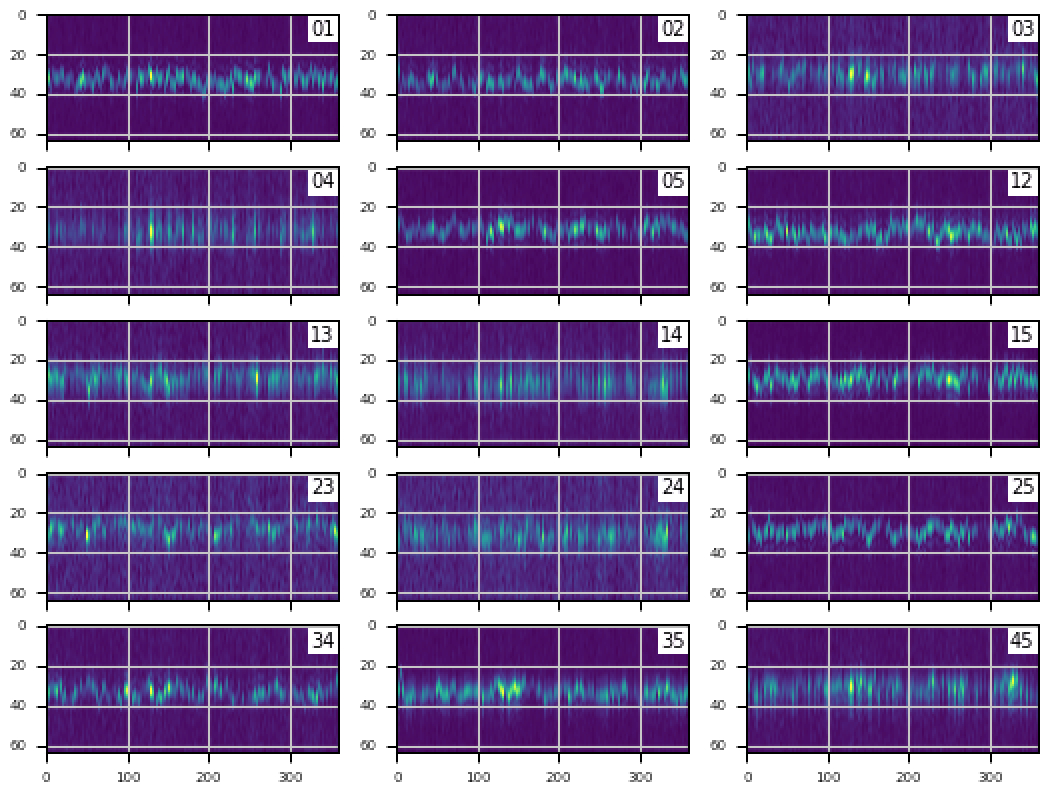} 
\caption{ A waterfall plot can be used to evaluate quickly the group delay tracking performance and quality of the data. }
\label{fig:waterfall}
\end{figure}

We achieved first 6-telescope fringes with the new detector system on the 11th of July, 2017,
which, to our knowledge, marks the first time that a SAPHIRA detector is used as primary science 
detector for an interferometric instrument.
Already shortly after the commissioning run, the new detector was used for routine science observations.
Following standard data cosmetics routines, we find that the data can be realiable reduced and calibrated,
as illustrated in Fig.~\ref{fig:transff}, which shows a typical visibility transfer function.
Fig.~\ref{fig:iotapeg} shows visibility amplitudes and closure phases that were measured on the well-known 
binary system $\iota$\,Peg, with the expected strong visibility/phase modulations.
This data has been reduced with the preliminary version of the MIRC-X data reduction pipeline that is available at:
\url{https://gitlab.chara.gsu.edu/lebouquj/mircx_pipeline}.

During the data recording, it is also possible to monitor the data quality on the 15  baselines
and the group delay tracking performance using a waterfall representation, as shown in Fig.~\ref{fig:waterfall}.

\subsection{Remote operation}

One of the objectives of our software work was to enable remote observing capabilities
that would allow us to operate MIRC-X from dedicated remote observing stations
at the University of Exeter, University of Michigan and Georgia State University.

This operation mode is enabled by the new server-client scheme, where all real-time
operations and commands are executed by the MIRC-X server that runs on the on-site instrument computer.
On the remote machine, the CHARA software libraries and MIRC-X software need to be installed
and the observer runs the Linux GTK GUIs that communicate to the server via ssh-tunnelling using the remote port forwarding protocol. 

The GUIs request realtime information about the status of the instrument from the server and display this
information as well as the real-time data.
Using the GUIs, the observer can send commands to the server or request configuration changes
that are then executed on the server site.  
This communication protocol also permits us to open multiple instances of the operation GUIs on multiple sites
(e.g.\ to enable real-time monitoring of ongoing observations) or to restart GUIs without affecting or interrupting the real-time operation.

The remote observing software capabilities have been tested successfully from the University of Exeter in May 2018
and from the University of Michigan and Georgia State University in June 2018.

\subsection{MIRC-X / MYSTIC parallel operation}
\label{sec:MYSTICop}

MIRC-X and MYSTIC will cover complementary spectral bands, namely the J and H-band (MIRC-X) 
and the K-band (MYSTIC).  Covering such a wide wavelength range opens exciting opportunities, for instance, 
to obtain a robust estimate for the temperature structure in protoplanetary disks.
Running the existing group delay tracking algorithm on either instrument will then enable data recording
with the sibling instrument, which can be useful in cases where the target is significantly easier to track in one band.
For this purpose, the internal delay between MIRC-X and MYSTIC will be adjusted at the beginning of the night using differential delay lines
that we will be implemented with motorized linear stages.

In addition, we plan to implement a ``cross-instrument'' fringe tracking algorithm that will
use fringe detection on either instrument for group delay tracking.
This could enable observations on very resolved objects (such as YSOs), where the visibility on some baselines 
might be too low in one band, but sufficient in another band.

Furthermore, with a phase tracking algorithm running on one of the instruments, we should also be
able to enable spectro-interferometric observations with high spectral dispersion in the other instrument,
e.g.\ with MYSTIC acting as phase tracker, while MIRC-X records data in the bright J-band accretion and outflow-tracing spectral lines.
This mode will require no further hardware, but an optimised fringe phase tracking algorithm.

\section{CONCLUSIONS}
\label{sec:conclusions}

As part of Phase~1 of the MIRC-X instrumentation project, we installed and commissioned an
eAPD SAPHIRA-based C-RED One camera at the CHARA array.
We could confirm that the detector provides an order-of-magnitude improvement in sensitivity
compared to MIRC's earlier PICNIC detector, in particular when using short integration times. 
For integration times above 20 millisecond, the sensitivity gain is less pronounced, as the thermal background 
becomes the dominant noise source. 
The sensitivity gain observed on sky are somewhat lower, in line with some remaining known optical issues,
such as the present oversampling of the fringes, and a poor light injection in the MIRC-X fibers.
We plan to address many of these issues during an upcoming engineering run in September 2018,
when we will also implement further sensitivity and precision-enhacing features.\\

Further information and the latest news on our instrumentation work can be found at \url{http://mircx.skraus.eu}.

\acknowledgments     

MIRC-X is funded by a Starting Grant from the European Research Council (ERC; grant agreement No.\ 639889, PI: Kraus) and funds from the University of Exeter. 
The project builds on earlier investments from the University of Michigan and the National Science Foundation (NSF, PI: Monnier).
We thanks the CHARA staff for their excellent support during our commissioning work and operation.
This research has made use of the Jean-Marie Mariotti Center \texttt{OIFits Explorer} service (available at http://www.jmmc.fr/oifitsexplorer).


\bibliography{MIRCX}   

\begin{thebibliography}{10}

\bibitem{mon04}
J.~D. {Monnier}, J.-P. {Berger}, R.~{Millan-Gabet}, and T.~A. {ten Brummelaar},
  ``{The Michigan Infrared Combiner (MIRC): IR imaging with the CHARA Array},''
  in {\em New Frontiers in Stellar Interferometry},  W.~A. {Traub}, ed., {\em
  \procspie} {\bf 5491}, p.~1370, Oct. 2004.

\bibitem{mon06}
J.~D. {Monnier}, E.~{Pedretti}, N.~{Thureau}, J.-P. {Berger},
  R.~{Millan-Gabet}, T.~{ten Brummelaar}, H.~{McAlister}, J.~{Sturmann},
  L.~{Sturmann}, P.~{Muirhead}, A.~{Tannirkulam}, S.~{Webster}, and M.~{Zhao},
  ``{Michigan Infrared Combiner (MIRC): commissioning results at the CHARA
  Array},'' in {\em Society of Photo-Optical Instrumentation Engineers (SPIE)
  Conference Series},  {\em \procspie} {\bf 6268}, p.~62681P, June 2006.

\bibitem{mon10}
J.~D. {Monnier}, M.~{Anderson}, F.~{Baron}, D.~H. {Berger}, X.~{Che},
  T.~{Eckhause}, S.~{Kraus}, E.~{Pedretti}, N.~{Thureau}, R.~{Millan-Gabet},
  T.~{ten Brummelaar}, P.~{Irwin}, and M.~{Zhao}, ``{MI-6: Michigan
  interferometry with six telescopes},'' in {\em Optical and Infrared
  Interferometry II},  {\em \procspie} {\bf 7734}, p.~77340G, July 2010.

\bibitem{Roettenbacher2016}
R.~M. {Roettenbacher}, J.~D. {Monnier}, H.~{Korhonen}, A.~N. {Aarnio},
  F.~{Baron}, X.~{Che}, R.~O. {Harmon}, Z.~{K{\H o}v{\'a}ri}, S.~{Kraus}, G.~H.
  {Schaefer}, G.~{Torres}, M.~{Zhao}, T.~A. {Ten Brummelaar}, J.~{Sturmann},
  and L.~{Sturmann}, ``{No Sun-like dynamo on the active star {$\zeta$}
  Andromedae from starspot asymmetry},'' {\em \nat}~{\bf 533}, pp.~217--220,
  May 2016.

\bibitem{klo10}
B.~{Kloppenborg}, R.~{Stencel}, J.~D. {Monnier}, G.~{Schaefer}, M.~{Zhao},
  F.~{Baron}, H.~{McAlister}, T.~{ten Brummelaar}, X.~{Che}, C.~{Farrington},
  E.~{Pedretti}, P.~J. {Sallave-Goldfinger}, J.~{Sturmann}, L.~{Sturmann},
  N.~{Thureau}, N.~{Turner}, and S.~M. {Carroll}, ``{Infrared images of the
  transiting disk in the {$\epsilon$} Aurigae system},'' {\em \nat}~{\bf 464},
  pp.~870--872, Apr. 2010.

\bibitem{sch14}
G.~H. {Schaefer}, T.~T. {Brummelaar}, D.~R. {Gies}, C.~D. {Farrington},
  B.~{Kloppenborg}, O.~{Chesneau}, J.~D. {Monnier}, S.~T. {Ridgway},
  N.~{Scott}, I.~{Tallon-Bosc}, H.~A. {McAlister}, T.~{Boyajian}, V.~{Maestro},
  D.~{Mourard}, A.~{Meilland}, N.~{Nardetto}, P.~{Stee}, J.~{Sturmann},
  N.~{Vargas}, F.~{Baron}, M.~{Ireland}, E.~K. {Baines}, X.~{Che}, J.~{Jones},
  N.~D. {Richardson}, R.~M. {Roettenbacher}, L.~{Sturmann}, N.~H. {Turner},
  P.~{Tuthill}, G.~{van Belle}, K.~{von Braun}, R.~T. {Zavala}, D.~P.~K.
  {Banerjee}, N.~M. {Ashok}, V.~{Joshi}, J.~{Becker}, and P.~S. {Muirhead},
  ``{The expanding fireball of Nova Delphini 2013},'' {\em \nat}~{\bf 515},
  pp.~234--236, Nov. 2014.

\bibitem{mon18}
J.~{Monnier}, J.-B. {Le Bouquin}, N.~{Anugu}, S.~{Kraus}, B.~{Setterholm},
  J.~{Ennis}, C.~{Lanthermann}, L.~{Jocou}, and T.~{ten Brummelaar}, ``{MYSTIC:
  Michigan Young STar Imager at CHARA},'' in {\em Optical and Infrared
  Interferometry VI},  {\em \procspie}, July 2018.

\bibitem{dul10}
C.~P. {Dullemond} and J.~D. {Monnier}, ``{The Inner Regions of Protoplanetary
  Disks},'' {\em \araa}~{\bf 48}, pp.~205--239, Sept. 2010.

\bibitem{zha11}
M.~{Zhao}, J.~D. {Monnier}, X.~{Che}, E.~{Pedretti}, N.~{Thureau},
  G.~{Schaefer}, T.~{ten Brummelaar}, A.~{M{\'e}rand}, S.~T. {Ridgway},
  H.~{McAlister}, N.~{Turner}, J.~{Sturmann}, L.~{Sturmann}, P.~J.
  {Goldfinger}, and C.~{Farrington}, ``{Toward Direct Detection of Hot Jupiters
  with Precision Closure Phase: Calibration Studies and First Results from the
  CHARA Array},'' {\em \pasp}~{\bf 123}, p.~964, Aug. 2011.

\bibitem{gar18}
T.~{Gardner}, J.~D. {Monnier}, F.~C. {Fekel}, M.~{Williamson}, D.~K. {Duncan},
  T.~R. {White}, M.~{Ireland}, F.~C. {Adams}, T.~{Barman}, F.~{Baron}, T.~{ten
  Brummelaar}, X.~{Che}, D.~{Huber}, S.~{Kraus}, R.~M. {Roettenbacher},
  G.~{Schaefer}, J.~{Sturmann}, L.~{Sturmann}, S.~J. {Swihart}, and M.~{Zhao},
  ``{Precision Orbit of {$\delta$} Delphini and Prospects for Astrometric
  Detection of Exoplanets},'' {\em \apj}~{\bf 855}, p.~1, Mar. 2018.

\bibitem{kur16}
R.~{Kurosawa}, A.~{Kreplin}, G.~{Weigelt}, A.~{Natta}, M.~{Benisty},
  A.~{Isella}, E.~{Tatulli}, F.~{Massi}, L.~{Testi}, S.~{Kraus}, G.~{Duvert},
  R.~G. {Petrov}, and P.~{Stee}, ``{Probing the wind-launching regions of the
  Herbig Be star HD 58647 with high spectral resolution interferometry},'' {\em
  \mnras}~{\bf 457}, pp.~2236--2251, Apr. 2016.

\bibitem{kra12c}
S.~{Kraus}, J.~D. {Monnier}, X.~{Che}, G.~{Schaefer}, Y.~{Touhami}, D.~R.
  {Gies}, J.~P. {Aufdenberg}, F.~{Baron}, N.~{Thureau}, T.~A. {ten Brummelaar},
  H.~A. {McAlister}, N.~H. {Turner}, J.~{Sturmann}, and L.~{Sturmann}, ``{Gas
  Distribution, Kinematics, and Excitation Structure in the Disks around the
  Classical Be Stars {$\beta$} Canis Minoris and {$\zeta$} Tauri},'' {\em
  \apj}~{\bf 744}, p.~19, Jan. 2012.

\bibitem{eis14}
J.~A. {Eisner}, L.~A. {Hillenbrand}, and J.~M. {Stone}, ``{Constraining the
  sub-au-scale distribution of hydrogen and carbon monoxide gas around young
  stars with the Keck Interferometer},'' {\em \mnras}~{\bf 443},
  pp.~1916--1945, Sept. 2014.

\bibitem{credone2016}
J.-L. {Gach}, P.~{Feautrier}, E.~{Stadler}, T.~{Greffe}, F.~{Clop},
  S.~{Lemarchand}, T.~{Carmignani}, D.~{Boutolleau}, and I.~{Baker}, ``{C-RED
  one: ultra-high speed wavefront sensing in the infrared made possible},'' in
  {\em Adaptive Optics Systems V},  {\em \procspie} {\bf 9909}, p.~990913, July
  2016.

\bibitem{finger2014}
G.~{Finger}, I.~{Baker}, D.~{Alvarez}, D.~{Ives}, L.~{Mehrgan}, M.~{Meyer},
  J.~{Stegmeier}, and H.~J. {Weller}, ``{SAPHIRA detector for infrared
  wavefront sensing},'' in {\em Adaptive Optics Systems IV},  {\em \procspie}
  {\bf 9148}, p.~914817, Aug. 2014.

\bibitem{lan18}
C.~{Lanthermann}, J.-B. {Le Bouquin}, N.~{Anugu}, J.~D. {Monnier}, and
  S.~{Kraus}, ``{Astronomical interferometry with near-IR e-APD at CHARA:
  characterization, optimization and on-sky operation},'' in {\em Optical and
  Infrared Interferometry VI},  {\em \procspie}, July 2018.

\bibitem{anu18}
N.~{Anugu}, J.-B. {Le Bouquin}, J.~D. {Monnier}, S.~{Kraus}, J.~{Ennis},
  C.~{Lanthermann}, B.~{Setterholm}, C.~{Davies,}, T.~{ten Brummelaar},
  M.~{Haidar}, V.~{Dubravec}, and S.~{Peters}, ``{MIRC-X/CHARA: sensitivity
  improvements with an ultra-low noise SAPHIRA detector, instrument
  specifications and operation},'' in {\em Optical and Infrared Interferometry
  VI},  {\em \procspie}, July 2018.

\bibitem{lazareff2012}
B.~{Lazareff}, J.-B. {Le Bouquin}, and J.-P. {Berger}, ``{A novel technique to
  control differential birefringence in optical interferometers. Demonstration
  on the PIONIER-VLTI instrument},'' {\em \aap}~{\bf 543}, p.~A31, July 2012.

\end{thebibliography}
\bibliographystyle{spiebib}   

\end{document}